\documentclass[reprint,a4paper,aps,pra,amsmath,amsfonts,showpacs,floatfix,nofootinbib]{revtex4-1}
\pdfoutput=1
\usepackage{bm}                      % Bold Greek letters
\usepackage{graphicx}                % Extended graphics package
\usepackage{dcolumn}                 % Aligning table columns
\usepackage[table]{xcolor}
\usepackage{amsmath}
\usepackage{times}
\usepackage{braket}
\usepackage{float}
\usepackage{placeins}
\usepackage{url}
\PassOptionsToPackage{hyphens}{url}\usepackage{hyperref}

\allowdisplaybreaks

\newcommand*{\centt}[1]{\multicolumn{1}{c}{#1}}
\newcommand*{\cent}[1]{\multicolumn{1}{c}{$#1$}}
\newcolumntype{x}[1]{D{.}{.}{#1}}
\newcommand{\ds}{\displaystyle}

\newcommand{\icm}{\text{cm}^{-1}}

\newcommand{\SE}{Schr{\"o}dinger equation}
\newcommand{\Eel}{\mathcal{E}_\mathrm{el}}
\newcommand{\Ea}{\mathcal{E}_{\mathrm{a}}}
\newcommand{\Hel}{H_\mathrm{el}}
\newcommand{\Hn}{H_\mathrm{n}}

\newcommand{\Hinvp}{\frac{1}{(\Eel-\Hel)'}}
\newcommand{\me}{m_{\mathrm{e}}}
\newcommand{\mun}{\mu_{\mathrm{n}}}
\newcommand{\suma}{\sum_{a}}
\newcommand{\na}{\vec{\nabla}_{\!a}}
\newcommand{\nr}{\vec{\nabla}_{\!R}}
\newcommand{\nrs}{{\nabla}_{\!R}^2}
\newcommand{\nel}{\vec{\nabla}_{\!\mathrm{el}}}

\newcommand{\phiel}{\phi_\mathrm{el}}
\newcommand{\phina}{\delta\phi_\mathrm{na}}
\newcommand{\br}{\vec{r}}
\newcommand{\bR}{\vec{R}}
\newcommand{\subel}{{_\mathrm{el}}}
\newcommand{\cY}{\mathcal{Y}}
\newcommand{\cWII}{\mathcal{W}_{\parallel}}
\newcommand{\cWT}{\mathcal{W}_{\perp}}
\newcommand{\cU}{\mathcal{U}}
\newcommand{\cV}{\mathcal{V}}
\newcommand{\dV}{\delta\mathcal{V}}
\newcommand{\dEna}{\delta\mathcal{E}_{\mathrm{na}}}
\newcommand{\EVS}[1]{\left\langle\phiel\left|#1\right|\phiel\right\rangle_{\!\mathrm{el}}}

\begin{document}
\preprint{Version 2.0}

\title{Rovibrational energy levels of the hydrogen molecule
       through nonadiabatic perturbation theory}

\author{Jacek Komasa}
\affiliation{Faculty of Chemistry, Adam Mickiewicz University, Uniwersytetu Pozna\'nskiego 8, 61-614 Pozna{\'n}, Poland}

\author{Mariusz Puchalski}
\affiliation{Faculty of Chemistry, Adam Mickiewicz University, Uniwersytetu Pozna\'nskiego 8, 61-614 Pozna{\'n}, Poland}

\author{Pawe\l\ Czachorowski}
\affiliation{Faculty of Physics, University of Warsaw, Pasteura 5, 02-093 Warsaw, Poland}

\author{Grzegorz {\L}ach}
\affiliation{Faculty of Physics, University of Warsaw, Pasteura 5, 02-093 Warsaw, Poland}

\author{Krzysztof Pachucki}
\affiliation{Faculty of Physics, University of Warsaw, Pasteura 5, 02-093 Warsaw, Poland}

\date{\today}

\begin{abstract}
We present an accurate theoretical determination of rovibrational energy levels of the hydrogen
molecule and its isotopologues in its electronic ground state.
We consider all significant corrections to the Born-Oppenheimer approximation, obtained
within nonadiabatic perturbation theory, including the mixed nonadiabatic-relativistic
effects. Quantum electrodynamic  corrections in the leading $\alpha^5\,m$
and the next-to-leading $\alpha^6\,m$ orders, as well as finite nuclear size effect,
are also taken into account but within the Born-Oppenheimer approximation only.
Final results for the transition wavelength between rovibrational levels achieve
accuracy of the order of $10^{-3}$--$10^{-7}$ cm$^{-1}$, and are provided by simple to use computer code.
\end{abstract}
\maketitle

%\tableofcontents

\section{Introduction}

The hydrogen molecule is one of the simplest chemical systems.
Nevertheless, it has a rich spectrum of rovibrational levels with lifetimes 
of the order of $10^5$--$10^6$ s.
Thanks to these long lifetimes, the contemporary measurements of transition frequencies
between rovibrational levels in H$_2$ and its isotopologues has
reached an accuracy level of $10^{-8}$ and in unique cases even $10^{-9}$ 
\cite{Dickenson:13,Niu:14,Niu:15,Mondelain:16,Schloesser:17,
Altmann:18,Cheng:18,Cozijn:18,Fasci:18,Martinez:18b,Tao:18,Trivikram:18,Wcislo:18,Holsch:19}.
In parallel, recent advances in a theoretical method---the nonadiabatic perturbation theory 
(NAPT)---have enabled accurate prediction for an arbitrary vibration and rotation 
quantum number of an arbitrary hydrogen isotopologue. Thus, theoretical progress 
and experimental availability make it an attractive candidate for precise tests 
of quantum electrodynamic (QED) theory and the search for a new physics. 

This paper presents the framework of theoretical methods for
calculation of nonrelativistic energies and of relativistic and QED corrections
up to $\alpha^7\,m$ order, together with the improved calculations of the so called heteronuclear potential. 
All the recent advances, including the complete $\alpha^6\,m$ and the $\alpha^4\,m^2/M$ 
corrections, as well as direct four-body calculations of nonrelativistic energies,
have been included. Results of these calculations, in the form of pertinent
internuclear potentials have been implemented in a publicly available computer code~\cite{H2SPECTRE};
therefore, this work provides the highest accuracy of all the energy levels 
and transition frequencies for the hydrogen molecule in the ground electronic state.
Depending on the isotopic contents and on the molecular level,
the absolute accuracy ranges from $10^{-3}$ to $10^{-4}$ cm$^{-1}$,
mostly limited by the yet unknown $\alpha^5\,m^2/M$ correction
or the higher-order nonadiabatic nonrelativistic effects.

This work finds application in many areas of physics 
ranging from astrophysical observations \cite{Roueff:19}, 
through verification of experimental spectra \cite{Lai:19,Hussels:19}, 
to measurements of the neutrino mass~\cite{Bodine:15}.

\section{NRQED framework}

The most convenient theoretical framework for the accurate description of light molecular 
systems is nonrelativistic quantum electrodynamics (NRQED).
It is an effective theory derived from relativistic QED
by matching the scattering amplitude up to certain powers in external momenta \cite{Caswell:86}.
The main advantage of NRQED approach is that all the corrections are
implemented on the top of a nonrelativistic Hamiltonian, and one can
use the standard perturbation theory with the nonrelativistic wave function.

The principal assumption in NRQED, is that the total energy can be expanded in powers of the fine-structure 
constant $\alpha$ (with $m$ being the mass of the electron)
\begin{align}
  E(\alpha) =&\  \alpha^2m\,E^{(2)} + \alpha^4m\,\left(E^{(4)} + E^{(4)}_{\rm FS}\right) + \alpha^5m\,E^{(5)}
\nonumber \\ &\ 
  + \alpha^6m\,E^{(6)} + ..., \label{alphaseries}
\end{align}
where $E^{(4)}_{\rm FS}$ is the finite nuclear size correction.
$E^{(n)}$ may include also powers of $\ln\alpha$, which is not shown explicitly. 
The expansion terms are interpreted subsequently as the nonrelativistic energy $E^{(2)}$, 
the relativistic correction $E^{(4)}$, the leading QED correction $E^{(5)}$, 
and the higher-order QED corrections $E^{(i)},\,i\geq 6$.
All these contributions can be expressed as expectation values of certain operators derived within NRQED theory, with a Schr\"odinger wave function.

\section{Nonrelativistic energy $E^{(2)}$}

The leading term of Eq.~(\ref{alphaseries}) is an eigenvalue of 
the Coulomb-Schr\"odinger Hamiltonian (in atomic units) % $\alpha=\frac{e^2}{4\pi}$
\begin{align}
H^{(2)}&=-\sum_{a}\frac{\vec{\nabla}^{2}_{\!a}}{2\,m_a}+\sum_{a>b}\frac{Z_a\,Z_b}{r_{ab}},
\end{align} 
where $m_a$ is the mass of the $a$th particle, $Z_a$  its charge, and $a$, $b$ go over all 
of the particles in the system (two electrons and two nuclei).
This eigenvalue is determined using two distinct methods.
The first one, called here ``direct,'' relies on a variational solution of the four-body Schr{\"o}dinger
equation with fully nonadiabatic wave function expanded in the basis of the nonadiabatic
James-Coolidge (naJC) functions \cite{PK:16}. By this approach, the energy of a given rovibrational level 
can be evaluated to a very high accuracy but requires a separate computationally intensive variational calculation.
Hence, currently, a limited fraction of the energy levels evaluated with this method are available.
The second method relies on the expansion of the energy in a small parameter,
being a ratio of the electron-to-nuclei mass. Within this perturbative approach, called
the nonadiabatic perturbation theory (NAPT), a set of nuclear interaction potentials is constructed,
which, in turn, enable all the energy levels collectively to be determined in a single calculation
at insignificant computational cost. Both methods are briefly described in the following subsections.

\subsection{Direct evaluation of $E^{(2)}$}

This method is conceptually very simple. It relies on a variational solution
of the Schr{\"o}dinger equation
\begin{equation}\label{ESchrEq}
H^{(2)}\Psi=E^{(2)}\Psi
\end{equation}
by expanding the wave function $\Psi$ in a four-particle basis
\begin{eqnarray}\label{EPsi}
 \Psi &=& \sum_k c_k\, \psi_{\{k\}}(\vec r_A, \vec r_B, \vec r_1, \vec r_2)\,. %\\
% \psi_k &=& (1 + P_{1\leftrightarrow 2})\,(1+P_{A\leftrightarrow B})\,
% \phi_{\{k\}} (\vec r_A, \vec r_B, \vec r_1, \vec r_2)\,, 
 \end{eqnarray}
%where the $P_{i\leftrightarrow j}$ operator accounts for the symmetry with respect to
%the exchange of electrons (1, 2) or nuclei (A, B), which is applicable to homonuclear molecules.
The obtained matrix form of the Schr{\"o}dinger equation 
\begin{equation}
(\mathbb{H}-E^{(2)}\,\mathbb{S})\,\mathbb{C}=0
\end{equation}
is then solved using an inverse iteration method with efficiently parallelized linear algebra algorithms.
The basis functions employed in this expansion are the exponential functions of the form \cite{PK:16}
\begin{equation}
\psi_{\{k\}}=e^{-\alpha\,R-\beta(\zeta_1+\zeta_2)}\,
R^{k_0}\,r_{12}^{k_1}\,\eta_1^{k_2}\,\eta_2^{k_3}\,\zeta_1^{k_4}\,\zeta_2^{k_5},\label{EJCBasis}
\end{equation}
called nonadiabatic James-Coolidge (naJC) functions for their resemblance to the classic
functions introduced by James and Coolidge in 1933 \cite{James:33}.
The $R\equiv r_{AB}$ variable represents the internuclear distance, 
$r_{12}$ ---the inter\-electron distance and the meaning of the remaining variables 
present in this function is the following $\zeta_a=r_{aA}+r_{aB}$ and $\eta_a=r_{aA}-r_{aB}$.
The $\alpha$ and $\beta$ in Eq.~(\ref{EJCBasis}) denote 
nonlinear variational parameters, common for the whole set of basis functions called ``sector,'' 
and $k_j$ are non-negative integers collectively denoted as $\{k\}$. 
If needed, two or more sectors (with different pairs of $\alpha^{(i)}$ and $\beta^{(i)}$) can be used. 
Technical details concerning the evaluation of the matrix elements of $\mathbb{H}$ and 
$\mathbb{S}$ can be found in Refs. \cite{PK:16,PK:18a,PK:18b}.

 From the study of the energy convergence with the increasing size of the basis set,
we observe that the nonrelativistic results obtained with this method reach 
a relative numerical accuracy of $10^{-13}-10^{-14}$ \cite{PK:16,PK:18a,PK:18b,PK:19}. 
In principle, this accuracy can be further increased, but
in view of the fact that the uncertainties from physical constants 
appear at the level of $10^{-12}$, there is currently no need for such efforts.
All these available energies corresponding to particular molecules and rovibrational levels are
supplied to the computer code~\cite{H2SPECTRE} as external data to be read whenever needed.

\subsection{NAPT---the nonadiabatic perturbation theory}

NAPT was introduced in Refs. \cite{PK08b,PK09} and expanded later in
Refs. \cite{PK:10,PK:15}.
It relies on a perturbative separation of electronic and nuclear movements.
In the zeroth order, molecular energy is obtained from the Born-Oppenheimer (BO)
internuclear potential \cite{Pachucki:10b}. 
The next order, which goes with the ratio of the electron mass $\me$ 
to the nuclear reduced mass $\mu_n$, is represented as an adiabatic correction 
to the BO potential \cite{PK:14}. Higher-order corrections, which are quadratic in the mass ratio, 
give additional $R$-dependent potentials and $R$-dependent nuclear masses 
in the effective nuclear equation \cite{PK:15}. These higher-order nonadiabatic potentials 
lead to radically increased accuracy of nonrelativistic levels. 
This subsection provides a concise description of the NAPT method.

Let the total wave function $\Psi$ be a solution of the stationary Schr\"odinger 
equation~(\ref{ESchrEq})
with the Hamiltonian partitioned into the electronic and nuclear parts 
\begin{equation}
H^{(2)} = \Hel + \Hn\,.
\end{equation}
The clamped nuclei electronic Hamiltonian
\begin{equation}
\Hel = -\sum_{a}\frac{\nabla^2_a}{2\,\me} + V \label{EHel}
\end{equation}
consists of the electronic kinetic energy term and 
the potential $V$, which includes all the Coulomb interactions,
\begin{equation}
V = -\frac{1}{r_{1A}}-\frac{1}{r_{1B}}-\frac{1}{r_{2A}}-\frac{1}{r_{2B}}
+\frac{1}{r_{12}}+\frac{1}{R},
\end{equation}
with the fixed positions of the nuclei. % $\vec R_A$ and $\vec R_B$. 
After separation of the center of mass motion,
the nuclear Hamiltonian in the reference frame fixed at the geometrical center of the nuclei, is
\begin{eqnarray}
H_{\rm n}&=&- \frac{\nabla^2_{\!R}}{2\,\mu_{\rm n}}
           - \frac{{\nabla}_{\!\mathrm{el}}^2}{2\,\mu_{\rm n}}  \nonumber 
           - \biggl(\frac{1}{M_B}-\frac{1}{M_A}\biggr)\,
             \vec\nabla_R\cdot\nel\\ &=&
             H'_{\rm n}+H''_{\rm n}\,,\label{EHn}
\end{eqnarray}
where $\nel=\frac{1}{2}\suma\na$ and $\mun=\left(1/M_A+1/M_B\right)^{-1}$ 
is the nuclear reduced mass. The $H'_{\rm n}$ part is even with respect to the inversion
whereas the $H''_{\rm n}$ is odd and vanishes for a homonuclear molecule.
 
The unperturbed (zeroth-order) wave function is assumed in the form 
of the product
\begin{equation}
\Psi_{\rm a}(\br,\bR) = \phiel(\br;\bR) \; \chi(\bR) \label{Eadap}
\end{equation}
of the nuclear wave function $\chi$ and the electronic wave
function $\phiel$ which implicitly depends on the nuclear coordinates $\bR$.
The function $\phiel$ fulfills the electronic Schr\"odinger equation 
\begin{equation}
\Hel\,\phiel=\Eel(R)\,\phiel\,, \label{EelSch}
\end{equation}
while $\chi$ satisfies the nuclear equation with $\Eel(R)$ as an interaction potential. 

Having this in mind, the total wave function can be expressed as a sum of terms 
parallel to and orthogonal to $\phiel$
\begin{equation}
\Psi = \phiel\,\chi + \phina\,,
\end{equation}
where the latter means that the electronic matrix element
\begin{equation}
\langle\phina|\phiel\rangle\subel = 0 \label{Eortho}
\end{equation}
vanishes. The symbol $\langle\dots\rangle\subel$ used henceforth represents
an integration over electronic coordinates only.
With such a representation of wave function, the \SE~(\ref{ESchrEq})
can also be decomposed into parallel and orthogonal parts
\begin{equation}
\left[(\Hel-\Eel)+(\Eel+\Hn-E^{(2)})\right]\left|\phiel\,\chi+\phina\right\rangle = 0
\label{Edecomp}
\end{equation}
and rearranged further to
\begin{equation}
(\Eel-\Hel)|\phina\rangle = (\Eel+\Hn-E^{(2)})|\phiel\,\chi+\phina\rangle.
\end{equation}
Since $\phina$ is orthogonal to $\phiel$, 
the formal solution into the above equation can be expressed in the following recursive form
employing the reduced resolvent $\Hinvp$:
\begin{equation}
|\phina\rangle = \Hinvp
\bigl[ \Hn|\phiel\,\chi\rangle+(\Eel+\Hn-E^{(2)})|\phina\rangle\bigr]\label{Ephina}.
\end{equation}
The left multiplication of Eq.~(\ref{Edecomp}) by $\langle\phiel|$ gives
%\begin{equation}
%\langle\phiel|\Eel+\Hn-E
%|\phiel\,\chi+\phina\rangle_{\rm el} = 0\,,
%\end{equation}
%which can be rewritten to the form
\begin{equation}
(\Eel+\Ea+\Hn-E^{(2)})|\chi\rangle =
-\langle\phiel|\Hn|\phina\rangle_{\rm el}, 
\label{Erecur}
\end{equation}
with $\Ea(R)\equiv\langle\phiel|\Hn|\phiel\rangle_\mathrm{el}$ being the adiabatic
correction potential.
Finally, insertion of (\ref{Ephina}) to the above equation
forms a perturbative expansion for the effective nuclear Hamiltonian
\begin{equation}\label{Eeffective}
(\Eel+\Ea+\Hn-E^{(2)})|\chi\rangle =
-(\Hn^{(2)} + \Hn^{(3)} + \Hn^{(4)} + \ldots)|\chi\rangle.
\end{equation}
The leading terms of this series have the following explicit form:
\begin{align}
\Hn^{(2)} =& \biggl\langle\phiel\biggl|\Hn\,\Hinvp\,\Hn\,\biggr|\phiel\biggr\rangle\subel\,,\\
%\end{eqnarray}
%\begin{eqnarray}
\Hn^{(3)} =&\biggl\langle\phiel\biggl|\Hn\,\Hinvp\,(\Hn+\Eel-E^{(2)})\nonumber \\ &
\times\Hinvp\,\Hn\,\biggr|\phiel\biggr\rangle\subel\,,
%\end{eqnarray}
\intertext{and}
%\begin{eqnarray}
\Hn^{(4)} =&\biggl\langle\phiel\biggl|\Hn\,\Hinvp\,(\Hn+\Eel-E^{(2)})\,\Hinvp\nonumber \\ &
\times(\Hn+\Eel-E^{(2)})\,\Hinvp\,\Hn\,\biggr|\phiel\biggr\rangle\subel\,.
\end{align}

Let us concentrate for a while on homonuclear molecules ($\Hn''=0$).
Taking into account the form~(\ref{EHn}) of the nuclear Hamiltonian,
we can transform the above formulas further, e.g., 
\begin{eqnarray}\label{EHnt}
\Hn^{(2)} &=& 
\Bigl\langle \Hn\phiel\Bigl|\Hinvp\Bigr|\Hn\phiel\Bigr\rangle\subel\\&&
+\frac{1}{\mun}\,\nr\,\Bigl\langle\nr\phiel\Bigl|
\Hinvp\Bigr|
\Hn\phiel\Bigr\rangle\subel\nonumber\\&&
-\frac{1}{\mun}\,\Bigl\langle \Hn\phiel\Bigl|
\Hinvp\Bigr|
\nr\phiel\Bigr\rangle\subel\,\nr\nonumber \\&&
-\frac{1}{\mun^2}\,\nr\,\Bigl\langle\nr\phiel\Bigl|
\Hinvp\Bigr|
\nr\phiel\Bigr\rangle\subel\,\nr\,.\nonumber
\end{eqnarray}

The nuclear function $\chi(\bR)$ can be factorized into the product
of radial and angular parts, 
\begin{equation}
  \chi_{JM}(\vec R) = \frac{\chi_J(R)}{R}\,Y_{JM}(\vec n)\,,
\end{equation}
with spherical harmonic $Y_{JM}$ and $\vec{n}=\bR/R$. Such a factorization followed by integration
over the angular variables leads to the radial form of the nuclear Hamiltonian
%The second-order nonadiabatic effective Hamiltonian of Eq.~(\ref{EHnt})
%is transformed to the form \cite{PK08b,PK09}
\begin{eqnarray}\label{EHntwo}
\Hn^{(2)} &=& {\cal U}(R) + 
\biggl(\frac{2}{R} + \frac{\partial}{\partial R}\biggr)\,\cV(R)
\\ &&
-\frac{1}{R^2}\,\frac{\partial}{\partial R}\,R^2
\,\cWII(R)\,\frac{\partial}{\partial R}
+\frac{J\,(J+1)}{R^2}\,\cWT(R),
\nonumber
\end{eqnarray}
in which
\begin{align}
\cU(R) &= 
\Bigl\langle \Hn\phiel\Bigl|\Hinvp\Bigr|\Hn\phiel\Bigr\rangle\,,\\
\cV(R) &= \frac{1}{\mun}\,\Bigl\langle\partial_R\phiel\Bigl|
\Hinvp\Bigr|\Hn\phiel\Bigr\rangle\,,\\
\cWII(R) &=\frac{1}{\mun^2}\,\Bigl\langle\partial_R\phiel\Bigl|
\Hinvp\Bigr|\partial_R\phiel\Bigr\rangle\,,\\
\cWT(R) 
%&=\frac{1}{\mun^2}\,\frac{(\delta^{ij}-n^i\,n^j)}{2}\,
%\Bigl\langle\nabla_R^i\phiel\Bigl|\Hinvp\Bigr|
%\nabla_R^j\phiel\Bigr\rangle\nonumber \\ 
&=\frac{1}{2\,\mun^2\,R^2}\,\Bigl\langle \phiel\Bigl|\vec L_{\rm el}
\Hinvp\vec L_{\rm el}\Bigr|\phiel\Bigr\rangle\,.
\end{align}
In the last equation, it is assumed that the electronic wave function $\phi_{\rm el}$
in a ground molecular $\Sigma$ state is rotationally invariant,
which implies $\vec{L}_n=-\vec{L}_{\rm el}$ with $\vec L_n=-iR\times\nr$ and
$\vec L_{\rm el}=-i\sum_a \vec{r}_a\times\vec{\nabla_a}$\,.

The Hamiltonian $\Hn^{(2)}$ contains only the terms proportional to $(\me/\mun)^2$,
but $\Hn^{(3)}$ and $\Hn^{(4)}$ may have the terms with the second and higher powers 
of the electron-to-nucleus mass ratio. Because we are interested in the leading 
nonadiabatic correction, i.e. in the terms proportional to $(\me/\mun)^2$,
all the ${\cal O}((\me/\mun)^3)$ corrections  are neglected.
However, some terms from $\Hn^{(3)}$ and $\Hn^{(4)}$ can be represented as
$(\me/\mun)^2$ correction to the potential, or $(\me/\mun)^3$ to the nuclear kinetic energy.
Their representation is not unique and may have different but equivalent forms,
which differ by a commutator $[\Hn+{\cal E}_{\rm el} -E^{(2)} , Q]$ whose expectation value vanishes\
for an arbitrary $Q$.
These terms are also neglected for consistency reasons.
Hence we include only a term from the $\Hn^{(3)}$, which has a unique representation
and is the $(\me/\mun)^2$ correction to the potential \cite{PK09}, namely 
\begin{align}
\dV(R)&=-\frac{1}{2\,\mun^2}\,\partial_R\Eel\,\Bigl\langle\partial_R\phiel\Bigl|
\left[\Hinvp\right]^2\Bigr|
\partial_R\phiel\Bigr\rangle
\end{align}
and this correction is added to $\cV(R)$ in Eq~(\ref{EHntwo}). 
The omitted components of $\Hn^{(3)}$ and of the higher-order Hamiltonians
remain the main source of the uncertainty of the nonrelativistic results obtained within NAPT.
The magnitude of this uncertainty can be estimated for each level separately by the value
of the second-order NAPT correction to this level scaled by the $\me/\mun$ factor.
This estimation has been validated by the direct variational computations described 
in the preceding subsection.
Because the direct calculations give an energy that corresponds to the perturbative
series summed up to infinite order, the difference between the NAPT and the direct
results accounts for all the omitted higher-order corrections
\cite{PK:18a,PK:19}---their values turn out to be smaller than the simple 
uncertainty estimation by the scaling. It has also been found that
the missing contribution grows proportionally to $J(J+1)$ 
with a slope depending on the vibrational quantum number.
We should also mention that the higher-order nonadiabatic corrections are
singular at $R=0$. It means that the nonadiabatic expansion
does not work properly at a very small distance where the nuclear kinetic energy 
becomes comparable to the electronic one. Nevertheless, at $R=0$ the exact wave function 
is as small as $10^{-25}$ (in a.u.), which makes these singular terms numerically negligible,
but their existence indicates possible limitations of NAPT.

Equation~(\ref{Eeffective}), after reduction to one-dimensional form and the neglect
of the $\mathcal{O}((\me/\mun)^3)$ terms, can be explicitly written as \cite{PK09}
\begin{align}
&\biggl[-\frac{1}{R^2}\,\frac{\partial}{\partial R}\,
\frac{R^2}{2\,\mu_{\|}(R)}\,\frac{\partial}{\partial R}\,
+\frac{J\,(J+1)}{2\,\mu_\perp(R) R^2}\, 
+\mathcal{Y}(R)\biggr]\,\chi_J(R) \nonumber \\
&= E^{(2)}\,\chi_J(R) \,,\label{ENASE}
\end{align}
where the functions
\begin{equation}
\frac{1}{2\,\mu_{\|}(R)} \equiv \frac{1}{2\,\mu_{\rm n}} + {\mathcal W}_\|(R)
\label{ERdmu}
\end{equation}
and
\begin{equation}
\frac{1}{2\,\mu_\perp(R)}\equiv \frac{1}{2\,\mu_{\rm n}} + {\mathcal W}_\perp(R)
\label{ERdmi}
\end{equation}
can be interpreted as $R$-dependent vibrational and rotational masses, and where
the potential $\cY(R)$ for the movement of the nuclei consists of the BO potential $\Eel(R)$
\cite{Pachucki:10b}, 
the adiabatic correction $\Ea(R)$ \cite{PK:14}, and the nonadiabatic correction
$\delta {\cal E}_{\rm na}(R)$ \cite{PK:15} potentials. The latter correction is expressed in terms 
of the functions defined above:
\begin{equation}
\delta\mathcal{E}_\mathrm{na}(R) =
\mathcal{U}(R) +\biggl(\frac{2}{R}+\frac{\partial}{\partial R}\biggr)
[\mathcal{V}(R)+\delta\mathcal{V}(R)]\,. \label{dena}
\end{equation}

Let us now return to the heteronuclear case, i.e. to the $\Hn$ Hamiltonian~(\ref{EHn})
in its full form. The unitary transformation from Ref.~\cite{PK:10}
\begin{equation}
\tilde H = \left(e^{\lambda\,\br\cdot\nr}\right)^+\,H\,e^{\lambda\,\br\cdot\nr} \label{EUHU}
\end{equation}
with $\br=\suma\br_a$ and the nuclear mass asymmetry parameter
\begin{equation}\label{Elambda}
\lambda=-\frac{\me}{2}\,\biggl(\frac{1}{M_B}-\frac{1}{M_A}\biggr)
\end{equation}
enables the heteronuclear part of the potential to be expressed as
an additional correction to the potential only,
\begin{align}
\dEna'(R) &=\lambda^2\,\biggl[\EVS{\frac{1}{\me}\,\nrs
+\frac{1}{2}\,r^i\,r^j\,\nabla_R^i\,\nabla_R^j(V)}\nonumber\\
&+ \EVS{\br\cdot\nr(V)\,\frac{1}{(\Eel-H_{\rm el})'}\,\br\cdot\nr(V)}\biggr] \,, 
\label{EEnahetero}
\end{align}
so that, finally, 
\begin{equation}\label{EY}
\mathcal{Y}(R) = \Eel(R)+\Ea(R)+\dEna(R)+\dEna'(R) \,.
\end{equation}
There is, however, one problem with the heteronuclear
correction $\delta {\cal E}'_{\rm na}$, because it behaves for small $R$
as $1/R^4$ and thus is singular. As mentioned earlier in this paragraph, NAPT does not work
at very small nuclear distances, but this region is numerically insignificant.
In practice, one can modify this potential $\delta{\cal E}'_{\rm na}(R)=\delta{\cal E}'_{\rm na}(R')$
for $R<R'$ and check that for a small $R'$ the results do not depend on its choice at the aimed precision.

Often it is the dissociation energy $D_{vJ}$ of rovibrational levels
which is of interest. For this reason, we fix the origin of the energy scale 
to the separated atoms limit and make all the potentials vanish at infinity:
\begin{align}\label{EYint}
  \tilde\cY(R)&=\cY(R)-\cY(\infty)\,. 
\end{align}
Similarly, we subtract the asymptotic value %${\mathcal W}(\infty)=-\frac{\me}{4\mun^2}$ 
from ${\mathcal W}$ potentials, 
\begin{align}\label{EWint}
\tilde{\mathcal W}(R)&={\mathcal W}(R)-{\mathcal W}(\infty)\,,
\end{align}
so that the $R$-dependent mass functions~(\ref{ERdmu})
and~(\ref{ERdmi}) correctly tend to the reduced atomic mass.
In this convention, the eigenvalue $E^{(2)}$ of the Hamiltonian in Eq.~(\ref{ENASE})
corresponds to the negative of $D_{vJ}$.

Equation~(\ref{ENASE}) has been solved using two distinct numerical methods.
One based on the Numerov integration method \cite{Johnson:07} and the other
on the discrete variable representation (DVR) method, with mutual agreement
between both of them.
In this work we use the DVR method, due to its great efficiency and simplicity of implementation.

\subsection{Discrete Variable Representation}

The DVR method is a pseudospectral method, making use of both a discrete grid 
and an associated basis set. There are many different flavors of DVR, using various basis sets 
and crafted for different integration ranges. The variant employed here \cite{H2SPECTRE} 
rests on the Fourier-basis version proposed by Colbert and Miller 
in Ref.~\cite{Colbert:92}.
It assumes the following expansion of the radial nuclear wave function: 
\begin{align}
\chi(R)&=\sum_{n=1}^{N}f_n\phi_n(R),\label{dvreq1}
\end{align}
where $\phi_n(R)$ are particle-in-a-box functions
\begin{align}
\phi_n(R)&=\left(\frac{2}{b-a}\right)^{1/2}\sin\left[\frac{n\pi(R-a)}{b-a}\right],\label{dvrf1}
\end{align}
where $R\in[a,b]$.
The coefficients $f_n$ can be expressed via a numerical quadrature with weights $w_n$
\begin{align}
f_n&=\sum_{m=1}^{N}w_n\, \phi_n(R_m)\chi(R_m)\,.\label{dvreq2}
\end{align}
The position $R$ is discretized on $N$ points---equal to the number of basis functions $\phi_n$
\begin{align}
R_m&=a+\frac{m(b-a)}{N+1}, \enskip \Delta R=\frac{b-a}{N+1},
\end{align}
for $m=1,...,N$ (which means that neither $a$ nor $b$ are grid points themselves).
The weights $w_n$ are all equal to the grid separation $\Delta R$ in this type 
of DVR \cite{Groenenboom:93}.
Combining Eqs. (\ref{dvreq1}) and (\ref{dvreq2}), one gets
\begin{align}
\chi(R)&=\sum_{n=1}^{N}\sum_{m=1}^{N}\Delta R\,\phi_n(R_m)\chi(R_m)\phi_n(R)\nonumber\\
&=\sum_{m=1}^{N}\chi_m\, \varphi_m(R),
\end{align}
where $\varphi_m(R)$ is a DVR orthonormal position basis [not to be confused with 
the auxiliary basis of Eq.~(\ref{dvrf1})], and $\chi_m$ is proportional to the value 
of the wave function on the $R_m$ grid point
\begin{align}
\varphi_m(R)&=\sum_{n=1}^{N}\phi_n(R_m)\phi_n(R)\sqrt{\Delta R}\,,\label{dvrf2}\\
\chi_m&=\chi(R_m)\sqrt{\Delta R}\,.\label{dvrf3}
\end{align}
It can be shown \cite{Colbert:92} that for $\phi_n$ of Eq.~(\ref{dvrf1}) the DVR basis function $\varphi_n$
exhibits asymptotically [i.e. for $(a-b)\to\infty$ and $N\to\infty$] the following
property:
\begin{align}
\varphi_n(R_m)&=\frac{\delta_{nm}}{\sqrt{\Delta R}}\,.\label{dvreq4}
\end{align}
That is why the potential-energy matrices in the $\varphi_n$ DVR basis are diagonal:
\begin{align}
V_{ij}&=\sum_{n=1}^N\Delta R\, \varphi_i(R_n)V(R_n)\varphi_j(R_n)
 %=\sum_{n=1}^N\delta_{in}\delta_{jn}V(R_n)
 =\delta_{ij}V(R_j)
\end{align}
and
\begin{align}
\braket{\chi|V|\chi}&\approx\sum_{n=1}^N\sum_{i=1}^N\sum_{j=1}^N\Delta R\, \chi_i\chi_j V(R_n) \varphi_i(R_n)\varphi_j(R_n)\nonumber\\
%&=\sum_{n=1}^N\sum_{i=1}^N\sum_{j=1}^N \chi_i\chi_j V(R_n)\delta_{in}\delta_{jn}\nonumber \\
&=\sum_{n=1}^N\chi_n\chi_n V(R_n).\label{dvr_pot}
\end{align}

The interval of $R$ in our problem is $[0,\infty)$, so $a=0$.
In practical applications $b$ and $N$ cannot be infinite, but already 
values as small as $N=200$ and $R_N=10.0$ are usually sufficient for most of our purposes.
They only need to be increased when investigating highly excited vibrational states.

The matrix elements of differential operators are nondiagonal but still can
be expressed by simple formulas. Namely, the Hamiltonian matrix elements in the BO approximation 
are given by
\begin{align}
H_{ij}&=\ds\frac{1}{\mu_\mathrm{n}\Delta R^2}(-1)^{i-j}\left(\frac{1}{(i-j)^2}-\frac{1}{(i+j)^2}\right),\\
H_{ii}&=\ds\frac{1}{2\mu_\mathrm{n}\Delta R^2}\left(\frac{\pi^2}{3}-\frac{1}{2i^2}\right)+\frac{J(J+1)}{2\mu_\mathrm{n}R_i^2}+\Eel(R_i).
\label{dvr_bo}
\end{align}
%\begin{widetext}
%\begin{align}
%H_{ij}&=\left\{\begin{array}{lr}
%\ds\frac{1}{2\mu_\mathrm{n}\Delta R^2}\left(\frac{\pi^2}{3}-\frac{1}{2i^2}\right)+\frac{J(J+1)}{2\mu_\mathrm{n}R_i^2}+\Eel(R_i),&\text{for } i=j\\[2ex]
%\ds\frac{1}{\mu_\mathrm{n}\Delta R^2}(-1)^{i-j}\left(\frac{1}{(i-j)^2}-\frac{1}{(i+j)^2}\right),&\text{for } i\neq j
%\end{array}\right\}.\label{dvr_bo}
%\end{align}
%\end{widetext}
The nonadiabatic \SE\ (\ref{ENASE}) leads to a more elaborate 
formula---not only because of the ``distance-dependent masses'' 
%$\mathcal{W}_{\parallel}(R)$ and $\mathcal{W}_{\perp}(R)$ 
present, but also because $\mathcal{W}_{\parallel}(R)$ 
is subjected to differentiation:
\begin{widetext}
\begin{align}
H_{ij}= 
\left\{\begin{array}{lr}
\ds\frac{1}{\Delta R^2}\left(\frac{1}{2\mu_\mathrm{a}}+\mathcal{\tilde W}_{\parallel}(R_i)\right)
\left(\frac{\pi^2}{3}-\frac{1}{2i^2}\right)
\\  \hspace*{3ex}
\ds+\frac{\mathcal{\tilde W}_{\parallel}^{\prime}(R_i)}{R_i}
 +\frac{1}{2}\mathcal{\tilde W}_{\parallel}^{\prime\prime}(R_i)
 +\left(\frac{1}{2\mu_\mathrm{a}}+\mathcal{\tilde W}_{\perp}(R_i)\right)\frac{J(J+1)}{R_i^2}+\mathcal{\tilde Y}(R_i) &\text{for } i=j,\\[2ex]
\ds\frac{(-1)^{i-j}}{\Delta R^2}\left(\frac{1}{\mu_\mathrm{a}}+\mathcal{\tilde W}_{\parallel}(R_i)+\mathcal{\tilde W}_{\parallel}(R_j)\right)\left(\frac{1}{(i-j)^2}-\frac{1}{(i+j)^2}\right) &\text{for } i\neq j,
\end{array}\right.\label{dvr_napt}
\end{align}
\end{widetext}
where $\mathcal{\tilde W}_{\parallel}'$ and $\mathcal{\tilde W}_{\parallel}''$ are the first and second derivatives
of $\mathcal{\tilde W}_{\parallel}$ with respect to $R$, 
and $\mathcal{\tilde Y}(R)$ has been defined in Eqs.~(\ref{EY}) and (\ref{EYint}).
Note that in the above formula the reduced nuclear mass $\mu_\mathrm{n}$ is replaced with the reduced atomic mass $\mu_\mathrm{a}$
\begin{align}
\frac{1}{\mu_\mathrm{a}}&=\frac{1}{m_A+\me}+\frac{1}{m_B+\me}.
\end{align}
This is because  Eq. (\ref{dvr_napt}) is written with respect to the dissociation limit and, 
as discussed in Refs. \cite{PK09,PK:10}, $\mu_\perp(R)$ 
and $\mu_\parallel(R)$ tend to $\mu_\mathrm{a}$ 
for $R\rightarrow\infty$.

The accuracy of $\mathcal{W}_{\parallel}$, $\mathcal{W}_{\perp}$, 
and $\delta\mathcal{E}_\mathrm{na}$ is considered to be high enough to
not contribute to the total nonrelativistic uncertainty. This uncertainty 
is dominated by the estimate of the neglected NAPT term,
which is $m_\mathrm{e}/\mu_\mathrm{n}$ times the leading nonadiabatic correction.
This missing-term uncertainty tends to be the largest source of the total theoretical error
if the direct nonadiabatic results are not available.

In the case of the heteronuclear-specific correction $\delta\mathcal{E}'_\mathrm{na}$,
the former calculations \cite{PK:10} have been significantly improved here using 
the previously optimized ECG functions from Ref.~\cite{CPKP:18}.
As a result, the numerical uncertainty of $\delta\mathcal{E}'_\mathrm{na}$ is also negligible.

\section{Relativistic correction $E^{(4)}$}

The second term in the $\alpha$ expansion~(\ref{alphaseries})
is the leading relativistic correction of the order $\alpha^4m$.
It is the expectation value of the Breit-Pauli Hamiltonian
\begin{equation}
E^{(4)} = \langle \Psi | H^{(4)}| \Psi \rangle.
\label{ErelFirst}
\end{equation}
For the hydrogen molecule in the $^1\Sigma^+_g$ state (in which all the electron spin-dependent terms vanish)
this Hamiltonian takes the following form \cite{PSKP:18}:
\begin{widetext}
\begin{align} \label{BP}
H^{(4)}=&\ -\sum_a\frac{p^4_a}{8}-\sum_X\frac{p^4_X}{8m_X^3}
+\frac{1}{2}\sum_{a,X}\frac{1}{m_X}p_a^i\left(\frac{\delta^{ij}}{r_{aX}}+\frac{r_{aX}^ir_{aX}^j}{r^3_{aX}}\right)p_X^j
-\frac{1}{2}\,p_1^i\left(\frac{\delta^{ij}}{r_{12}}+\frac{r_{12}^ir_{12}^j}{r^3_{12}}\right)p_2^j \nonumber \\ &\
-\frac{1}{2\,m_A\,m_B}p_A^i\left(\frac{\delta^{ij}}{r_{AB}}+\frac{r_{AB}^ir_{AB}^j}{r^3_{AB}}\right)p_B^j
+\frac{\pi}{2}\sum_{a,X}\left(1+\frac{\delta_{s_X}}{m_X^2}\right)\delta^3(r_{aX})+\pi\delta^3(r_{12}),
\end{align}
\end{widetext}
where $a$ goes over the electrons ($1$ and $2$) and $X$ -- over the nuclei ($A$ and $B$), 
and $\delta_{s}$ depends on the nuclear spin $s$: $\delta_{s}=0$ for $s=0$ or 1, 
and $\delta_{s}=1$ for $s=1/2$.
Its first two terms account for the relativistic correction to the kinetic energy.
The third, fourth, and fifth terms are called the Breit corrections (or ``orbit-orbit coupling'' terms) 
and can be attributed to the relativistic retardation of the Coulomb potential \cite{Bethe:57}.
The remaining contributions are represented by the so-called contact terms and are proportional
to the 3D Dirac delta functions.
In practice, even if the above Hamiltonian is used in the fully nonadiabatic approach 
(treating electrons and nuclei on an equal footing), the second term is 
neglected---being proportional to the very small $m_e^3/m_X^3$ factor.

For the dissociation energy $D$, one subtracts $E^{(4)}$ from the relativistic correction 
for separated atoms $E^{(4)}_A+E^{(4)}_B$, where
\begin{eqnarray}\label{Erelx}
  E^{(4)}_X &=& 
  -\frac{1}{8}\, + \frac{1}{4}\,\bigg(\frac{1}{m_X}\bigg)^2 + O\bigg(\frac{1}{m_X}\bigg)^3.
\end{eqnarray}
Note that the term proportional to $1/m_X$ is not present in the above formula; consequently,
the relativistic recoil correction for separated atoms is very small.

The direct calculations of the relativistic correction with the nonadiabatic wave function
have been performed only for the ground molecular state $v=0,J=0$ \cite{PSKP:18,PKSP:19}. 
The results for arbitrary vibrationally and rotationally excited states
have been obtained within the NAPT approach,
\begin{equation}
E^{(4)}=E^{(4,0)}+E^{(4,1)} +\ldots\,,
\end{equation}  
described in the following subsections.

\subsection{Leading-order relativistic correction, $E^{(4,0)}$}

The leading relativistic contribution in the BO approximation
consists of the nuclear-mass-independent terms from the Breit-Pauli Hamiltonian (\ref{BP}):
\begin{align}
H^{(4,0)}&=-\frac{p^4_1+p^4_2}{8} -\frac{1}{2}p_1^i\left(\frac{\delta^{ij}}{r_{12}}+\frac{r_{12}^ir_{12}^j}{r^3_{12}}\right)p_2^j +\pi \delta^3(r_{12}) \nonumber \\
&\quad+\frac{\pi}{2}\left(\delta^3(r_{1\mathrm{A}})+\delta^3(r_{2\mathrm{A}})+\delta^3(r_{1\mathrm{B}})+\delta^3(r_{2\mathrm{B}})\right).\label{eq350}
\end{align}
The correction to the BO potential energy is the expectation value with the electronic wave function
\begin{align}
\mathcal{E}^{(4,0)}(R)&=\braket{\phi_{\rm el}|H^{(4,0)}|\phi_{\rm el}}.\label{eq60}
\end{align}
It was observed in \cite{PKP:17} that the numerical convergence is significantly improved when the
electronic wave function $\phi_{\rm el}$ satisfies the electron-electron cusp condition.
The final value of the leading-order relativistic correction is evaluated as
the expectation value with the nuclear wave function
\begin{align}
E^{(4,0)}&=\braket{\chi|\mathcal{E}^{(4,0)}(R)|\chi} \label{eq61}
\end{align}
and this is calculated using the DVR function from the nonrelativistic BO approximation.
On the basis of the results of Ref.~\cite{PKP:17}, a numerical uncertainty $\delta E^{(4,0)}$
is estimated by $2\times10^{-6}\,E^{(4,0)}$.

\subsection{Finite nuclear mass relativistic correction $E^{(4,1)}$}

The leading finite-nuclear-mass relativistic correction can also be expressed 
in terms of effective internuclear potential and consists of three parts~\cite{CPKP:18},
\begin{align}
  \mathcal{E}^{(4,1)}(R)&= \mathcal{E}_1^{(4,1)}(R) + \mathcal{E}_2^{(4,1)}(R) + \mathcal{E}_3^{(4,1)}(R), \label{reladcompletef}
\end{align}
where
\begin{align}
\mathcal{E}_1^{(4,1)}(R)&=\frac{1}{\mu_\mathrm{n}}\braket{\vec{\nabla}_R\phi_\mathrm{rel}|\vec{\nabla}_R\phi_{\rm el}}, \label{corra1f} \\
\mathcal{E}_2^{(4,1)}(R)&=-\frac{1}{\mu_\mathrm{n}}\braket{\phi_\mathrm{rel}|\vec{\nabla}^2_\mathrm{el}|\phi_{\rm el}}, \label{corra2f} \\
\mathcal{E}_3^{(4,1)}(R)&= \braket{\phi_{\rm el}|H^{(4,1)}|\phi_{\rm el}},\label{corra3f}
\end{align}
and $\phi_\mathrm{rel}$ is a relativistic correction to the BO electronic wave function:
\begin{align}
\ket{\phi_\mathrm{rel}}&=\frac{1}{(\mathcal{E}\subel-H\subel)'}H^{(4,0)}\ket{\phi_{\rm el}}. \label{psirel}
\end{align}
The Hamiltonian $H^{(4,1)}$ describes the electron-nucleus Breit interaction, which
in the coordinate system assumed in this work takes the form
\begin{align}
&H^{(4,1)}=\label{hm}\\ & \nonumber 
              -\frac{1}{4\mu_\mathrm{n}}\sum_{a=1,2}\nabla_a^{i}\left(\frac{\delta^{ij}}{r_{a\mathrm{A}}}
               +\frac{r_{a\mathrm{A}}^ir_{a\mathrm{A}}^j}{r^3_{a\mathrm{A}}}-\frac{\delta^{ij}}{r_{a\mathrm{B}}}-\frac{r_{a\mathrm{B}}^ir_{a\mathrm{B}}^j}{r^3_{a\mathrm{B}}}\right)\nabla_R^{j}\\\nonumber
              &+\frac{1}{4\mu_\mathrm{n}}\sum_{a=1,2}\nabla_a^{i}\left(\frac{\delta^{ij}}{r_{a\mathrm{A}}}
               +\frac{r_{a\mathrm{A}}^ir_{a\mathrm{A}}^j}{r^3_{a\mathrm{A}}}+\frac{\delta^{ij}}{r_{a\mathrm{B}}}+\frac{r_{a\mathrm{B}}^ir_{a\mathrm{B}}^j}{r^3_{a\mathrm{B}}}\right)\nabla_\mathrm{el}^{j}.
\end{align}
This effective internuclear potential $\mathcal{E}^{(4,1)}$ is used to obtain the relativistic recoil
correction to rovibrational levels, using
\begin{align}
E^{(4,1)}=&\ \braket{\chi|\mathcal{E}^{(4,1)}(R)|\chi}\label{eq62}\\
&\ +2\braket{\chi|\mathcal{E}^{(4,0)}(R)\frac{1}{(E^{(2,0)}-H_{\rm n})'}\mathcal{E}^{(2,1)}(R)|\chi},\nonumber
\end{align}
where $\mathcal{E}^{(2,1)}(R)=\mathcal{E}_{\rm a}(R)$. The potential $\mathcal{E}^{(4,1)}(R)$ 
has been reported recently in Ref.~\cite{CPKP:18}.
The numerical error contributed by the potential was estimated to be 
${2\times10^{-4}\braket{\chi|\mathcal{E}^{(4,1)}|\chi}}$.
Furthermore, because currently no higher finite-nuclear-mass relativistic corrections are known, 
the effect of their omission is approximated by $E^{(4,1)} m_\mathrm{e}/\mu_\mathrm{n}$ 
and included in the total $E^{(4)}$ error estimate.

\section{QED corrections}

\subsection{Leading-order QED correction $E^{(5)}$}

The complete formula for the leading quantum electrodynamic correction $E^{(5)}$ for H$_2$ 
and its isotopologues was obtained in Refs.~\cite{PKCP:19,PKSP:19}.
Direct (four-body) numerical calculations have been performed only for the ground molecular 
level, whereas for all the excited levels we use the BO approximation.
The leading QED Born-Oppenheimer contribution can be expressed as
\begin{align}
E^{(5,0)}&=\braket{\chi|\mathcal{E}^{(5,0)}(R)|\chi},
\end{align}
where
\begin{align}
  \mathcal{E}^{(5,0)}(R)&=\frac{4}{3}\left[\frac{19}{30}-2\ln\alpha-\ln k_0(R)\right]
  \sum_{a,X}\braket{\delta^3(r_{aX})}_{\rm el}\nonumber\\
  &+\left[\frac{164}{15}+\frac{14}{3}\ln\alpha\right]\braket{\delta^3(r_{12})}_{\rm el}
  -\frac{7}{6\pi}\left\langle\frac{1}{r_{12}^3}\right\rangle_{\!\!{\rm el}, \epsilon}.\label{ma5pot}
\end{align}
In the above formula the expectation values are evaluated 
with the nonrelativistic wave function $\phi_{\rm el}$, and the notation 
$\langle 1/r_{ij}^3 \rangle_\epsilon$ means the following:
\begin{align}
\left\langle\frac{1}{r_{ij}^3}\right\rangle_{\!\!\epsilon} =
\lim_{\epsilon\rightarrow0}\left[\left\langle\frac{\theta(r_{ij}-\epsilon)}{r_{ij}^3}\right\rangle
+4\pi(\gamma+\ln \epsilon)\braket{\delta^3(r_{ij})}\right]\label{AS},
\end{align}
where the symbol $\gamma$ denotes the Euler-Mascheroni constant, and $\theta$ 
is the Heaviside function. The Bethe logarithm $\ln k_0(R)$ is
\begin{align}
  \ln k_0(R)&=\frac{\braket{\phiel|\,\vec{j}\,(H_\mathrm{el}-\mathcal{E}_\mathrm{el})
      \ln[2(H_\mathrm{el}-\mathcal{E}_\mathrm{el})]\,\vec{j}\,|\phiel}}
      {\braket{\phiel|\vec{j}(H\subel-\mathcal{E}\subel)\vec{j}|\phiel}},\label{logb}
\end{align}
with $ \vec{j}=-\vec{p}_1/\me-\vec{p}_2/\me$.
It has been calculated in Ref.~\cite{PLPKPJ09},
whereas the results for the Araki-Sucher term and Dirac $\delta$ are taken
from newer calculations reported in Refs. \cite{PKCP:16,PKP:17}. 
The numerical uncertainty is estimated to be ca.
$5\times 10^{-4} \braket{\chi|\mathcal{E}^{(5,0)}(R)|\chi}$.
The greatest source of error in this term comes from the uncalculated finite-nuclear-mass contribution,
estimated as $E^{(5,0)} m_\mathrm{e}/\mu_\mathrm{n}$.
For the levels and transitions where the nonrelativistic contribution is calculated directly
(so the NAPT error is removed), it dominates the total theoretical uncertainty.

\subsection{Higher-order QED correction $E^{(6)}$}

The higher-order QED contribution is calculated within the BO approximation 
and is given by
\begin{align}
\mathcal{E}^{(6,0)}(R)&=\braket{\phiel|H^{(6,0)}|\phiel} \nonumber\\
&+\braket{\phiel|H^{(4,0)}\frac{1}{(\mathcal{E}\subel-H\subel)'}H^{(4,0)}|\phiel}\label{ma6pot},
\end{align}
where $H^{(4,0)}$ is the Breit Hamiltonian in the nonrecoil limit,
and $H^{(6,0)}$ is the $O(\alpha^2)$ correction to this Hamiltonian. 
The explicit formulas for $\mathcal{E}^{(6,0)}(R)$ are far too extensive to be presented here.
They can be found in Ref. \cite{PKCP:16}. The total energy contribution in this order is
\begin{align}
E^{(6,0)}&=\braket{\chi|\mathcal{E}^{(6,0)}(R)|\chi} \nonumber\\
&+\bra{\chi}\mathcal{E}^{(4,0)}(R)\frac{1}{(E^{(2,0)}-H_{\rm n})'}\mathcal{E}^{(4,0)}(R)\ket{\chi}.
\end{align}
The second term in the above equation is again the second-order relativistic correction 
with respect to the relativistic BO potential (in our former works presented separately
as $E^{(6)}_\mathrm{sec}$ correction).
The $\mathcal{E}^{(6,0)}(R)$ potential was calculated in Ref.~\cite{PKCP:16}
and $\mathcal{E}^{(4,0)}(R)$ in Ref.~\cite{PKP:17}.
The numerical error was estimated as ${3\times 10^{-3} E^{(6,0)}}$,
whereas the missing finite-nuclear-mass correction was estimated as ${E^{(6,0)} m_\mathrm{e}/\mu_\mathrm{n}}$.

\subsection{Estimation of $E^{(7)}$}

The $E^{(7)}$ correction is of the highest order considered so far for the hydrogen molecule.
Currently, its complete form is unknown.
Here, we follow Ref.~\cite{PKCP:19} and include the leading
one- and two-loop radiative corrections known from the hydrogen atom (see \cite{Eides:01})
in the BO approximation
\begin{equation}
E^{(7)}=\braket{\chi|\mathcal{E}^{(7)}(R)|\chi},
\end{equation}
where
\begin{align}
&\mathcal{E}^{(7)}(R) \approx\  \pi\,\braket{\phiel|\sum_{a,X}\delta^3(r_{aX})|\phiel}\subel
  \bigg\{ \frac{1}{\pi}\big[ A_{60} \label{EE7}\\&\  + A_{61}\,\ln\alpha^{-2} 
  + A_{62}\, \ln^2\alpha^{-2}\big] + \frac{1}{\pi^2}\,B_{50} + \frac{1}{\pi^3}\,C_{40}\biggr\}.
\nonumber
\end{align}
As an uncertainty of the $E^{(7)}$ correction, following Ref.~\cite{PKCP:19}, we assume $25\%$ of its value.

\section{Finite nuclear size effect $E_\mathrm{FS}$}

At the achieved accuracy level, the nuclear finite size effect cannot be neglected anymore.
This correction, when evaluated in the BO approximation,
\begin{equation}
  E^{(4)}_{\rm FS}=\braket{\chi|\mathcal{E}^{(4)}_{\rm FS}(R)|\chi},
\end{equation}  
is accounted for by the following formula:
\begin{align}
\mathcal{E}^{(4)}_{\rm FS}(R) &=
\frac{2\pi}{3}\,\braket{\phiel|\sum_{a,X}\delta^3(r_{aX})|\phiel}\subel\,\frac{(r_{C,A}^2+r_{C,B}^2)}{2\,\lambdabar^2},
\label{FS}
\end{align}
where $\lambdabar$ 
is the reduced electron Compton wavelength;
$r^2_{C,X}$ is the mean square charge radius of the nucleus $X=A,B$, with
$r_p =  0.8414(19)$ fm \cite{CODATA:18}, $r_d = 2.12799(74)$ fm \cite{CODATA:18},
and $r_t =  1.7591(363)$ fm \cite{Angeli:13}.
Any higher-order effects due to the nuclear size 
or nuclear polarizability are neglected. 
The accuracy of $E^{(4)}_{\rm FS}$ is limited by the accuracy of the charge radii. 
The connection between the dissociation energy and the charge radius of a nucleus
can potentially be utilized to determine the latter one provided that both theoretical
and experimental dissociation energy are known to a sufficient accuracy, which is about $10^{-7}$ cm$^{-1}$.

\section{Uncertainty estimation}

Previous sections devoted to individual components $E^{(i)}$ 
of the $\alpha$ expansion~(\ref{alphaseries}) contain a short description 
of the uncertainty estimates $\delta E^{(i)}$, which are assumed to be uncorrelated.
Therefore, the total uncertainty is the square-root of the sum of squares of all the partial uncertainties.
Depending on the availability of the direct nonadiabatic results for a given 
rovibrational level we can distinguish three different cases.
In the first case, the direct nonadiabatic results are available for
the nonrelativistic energy as well as for relativistic and QED corrections.
In this case, currently represented by the ground levels of all the isotopologues, 
the dominating uncertainty comes from the incomplete knowledge of the $E^{(7)}$ term.
In the second case, only the nonrelativistic energy $E^{(2)}$ is known
with high accuracy from the direct nonadiabatic calculations. In such a case,
the overall accuracy is limited by the lack of the recoil correction to the leading
QED term $E^{(5)}$.
Finally, in the third and the most common case, all the energy components
are evaluated from the NAPT. Then, the limitations in accuracy originate either
from the nonrelativistic or QED component of the energy. 

The estimation of the uncertainty assigned to a transition energy is more complicated.
Depending on the pair of the states involved in a given transition, we observe
smaller or larger cancellation of different energy components. A systematic description
of this cancellation is difficult, and we assumed in general that the uncertainty assigned
the a transition energy is equal to the larger uncertainty out of these two states.
However, in particular cases, like the fundamental $\nu=0\rightarrow1$ transitions,
the cancellation of uncertainties is significant, and we associate relative $\me/\mun$
uncertainty to the energy difference, as demonstrated in Table~\ref{TTrans}. 

We also note that, because at the long-distance points the accuracy of the potentials 
usually deteriorates, our error estimates for highly excited levels can be inaccurate.

\section{Results and summary}

\begin{table*}[!t]
\caption{Selected transition energies (in $\icm$) obtained for H$_2$ from the NAPT
  and direct nonadiabatic calculations with breakdown into components.
  CODATA 2018 \cite{CODATA:18} values of physical constants are used. 
  For a shorthand notation, we shall identify the $\alpha^nm\,E^{(n)}$ terms with
  the bare coefficients $E^{(n)}$.}
\label{TTrans}
\begin{ruledtabular}
\begin{tabular*}{\textwidth}{l@{\extracolsep{\fill}}x{5.11}x{5.11}x{5.11}x{5.11}x{5.11}}
\rule{0pt}{2.3ex}
          & \cent{{(0,1)\text{ --- }(0,0)}} & \cent{{(1,0)\text{ --- }(0,0)}}& \cent{{(1,1)\text{ --- }(0,1)}} & \cent{{(2,1)\text{ --- }(0,3)}} &  \cent{(3,5)\text{ --- }(0,3)} \\
\hline\rule{0pt}{2.9ex}\!
$E^{(2)}$(NAPT)      &118.485\,262(7)    &4\,161.164\,2(9)      &4\,155.252\,0(9)      &7\,488.283\,3(17)     &12\,559.750\,0(25)\\
$E^{(2)}$(direct)    &118.485\,260\,5(1) &4\,161.164\,070\,0(1) &4\,155.251\,869\,3(1) &7\,488.283\,212\,0(1) &12\,559.749\,918\,5(1)  \\
$E^{(4)}$            &  0.002\,583\,6    &     0.023\,553\,9(2) &     0.023\,333\,3(2) &     0.028\,570\,7(3) &      0.065\,877\,6(6) \\
$E^{(5)}$            & -0.001\,022\,7(12)&    -0.021\,318(26)   &    -0.021\,257(25)   &    -0.036\,018(43)     &     -0.065\,815(79)  \\
$E^{(6)}$            & -0.000\,008\,9    &    -0.000\,191\,3(6) &    -0.000\,190\,8(6) &    -0.000\,326\,3(10)    &     -0.000\,594\,9(19) \\
$E^{(7)}$            &  0.000\,000\,5(1) &     0.000\,010\,3(26)&     0.000\,010\,3(26)&     0.000\,017\,4(44)    &      0.000\,031\,9(80) \\
$E_\mathrm{FS}^{(4)}$& -0.000\,000\,2    &    -0.000\,003\,2    &    -0.000\,003\,2    &    -0.000\,005\,4    &     -0.000\,009\,8   \\[1pt]
$E$                  &118.486\,812\,8(12)&4\,161.166\,122(26)   &4\,155.253\,762(26)   &7\,488.275\,451(43)     &12\,559.749\,408(79)  \\[1ex]
Exp.                 &118.486\,8(1)\ \text{\cite{Jennings:84}}&4\,161.166\,36(15)\ \text{\cite{Niu:14}}&4\,155.254\,00(21)\ \text{\cite{Niu:14}}&7\,488.275\,3(10)\ \text{\cite{Campargue:12}}&12\,559.749\,39(22)\ \text{\cite{Cheng:12}}\\
Diff.                & 0.000\,0(1)      &    -0.000\,24(15)    &    -0.000\,24(21)    &    +0.000\,2(10)    &     +0.000\,02(23)  \\
\end{tabular*}
\end{ruledtabular}
\end{table*}

One of the most pronounced features of NAPT combined with $\alpha$ expansion
in Eq. (\ref{alphaseries}) is that it gives access to an arbitrary
bound rovibrational energy level within the electronic ground state. Consequently,
it enables all transitions to be obtained within this manifold of levels.
Another merit of NAPT with $\alpha$ expansion is the possibility of full control of the accuracy of the results,
as well as the potential of gradually increasing this accuracy by improving the existing,
and adding new, terms to the expansion series.

The theoretical underpinning presented in this work as well as numerical calculations
performed over the past years enabled construction of a computer program \cite{H2SPECTRE}
serving the numerical values of the rovibrational energy levels and splittings
between them for all the isotopologues of the hydrogen molecule.
There are several thousands of such levels and many more transitions
available from this program. It would be impractical to present such
a large amount of data in printed form. Therefore,
this program has been made publicly available to the scientific community
so that numerical results for levels or transitions of interest
(in particular also for all of them) can be easily generated by the reader.
This form of the presentation of the results has also another important advantage---it
is our intention to support the program in the future by updating the input potentials
and physical constants and possibly by adding new functionalities---a guarantee that
it has the best currently available data.

Here, we present only a small selection of the numerical results to illustrate
the most important features of NAPT combined with $\alpha$ expansion.
In Table~\ref{TCmpr} we show the total dissociation energy $E$ for the ground level of
three lightest isotopologues. These energies are compared with the reference
theoretical results obtained from direct nonadiabatic calculations and with the best
available experimental data. This comparison shows the current accuracy limitations of NAPT
but simultaneously demonstrates that this method performs very well because its results agree
within uncertainties with direct variational calculations. 

\begin{table*}[!ht]
\caption{Comparison of dissociation energies (in $\icm$) of the ground levels of H$_2$, D$_2$, and HD
  obtained in the framework of NAPT with the results of direct nonadiabatic calculations 
  and with experimental data. CODATA 2018 \cite{CODATA:18} values of physical constants are used.}
\label{TCmpr}
\begin{ruledtabular}
\begin{tabular*}{\textwidth}{l@{\quad}*{3}{x{8.15}}}
\rule{0pt}{2.3ex}
          & \centt{H$_2$\qquad\qquad} & \centt{{D$_2$\qquad\qquad}} & \centt{{HD\qquad\qquad}}  \\
\hline\rule{0pt}{2.9ex}
\!$E$ (NAPT)         & 36\,118.069\,45(53)  &    36\,748.362\,27(17)     & 36\,405.782\,37(33)\\
$E$ (direct)       & 36\,118.069\,632(26)\ \text{\cite{PKCP:19}} &    36\,748.362\,342(26)\ \text{\cite{PKSP:19}} & 36\,405.782\,478(26)\ \text{\cite{PKSP:19}}   \\
Experiment         & 36\,118.069\,45(31) \ \text{\cite{Altmann:18}}   & 36\,748.362\,86(68) \ \text{\cite{Liu:09}}   & 36\,405.783\,66(36) \ \text{\cite{Sprecher:10}}  \\
%Diff.              &      -0.000\,18(31)     &       0.000\,57(68)     &       0.001\,18(36)    \\
\end{tabular*}
\end{ruledtabular}
\end{table*}

The finite nuclear mass effects are the most significant in H$_2$ because it is the lightest 
isotopologue. Therefore, Table~\ref{TTrans} contains a few examples of transitions 
between rovibrational levels of H$_2$ with growing energies.
In most cases, theoretical energies are an order of magnitude more accurate
than experimental values and in agreement with them.
However, we observe a significant  $3\,\sigma$ discrepancy between the measured dissociation energy
and our calculations for HD, in spite of a good agreement for H$_2$ and D$_2$; see Table II.
Before drawing any conclusions, this experimental value should be verified.

\begin{acknowledgments}
This research was supported by National Science Center (Poland) Grants No. 2016/23/B/ST4/01821  (M.P.)
and No. 2017/25/B/ST4/01024 (J.K.) as well as by a computing grant from 
Pozna\'n Supercomputing and Networking Center and by PL-Grid Infrastructure.
\end{acknowledgments}

%\bibliography{naH2,H2spectr} 

\end{document}